# Structure of nuclei at extreme values of the isospin

J. Dobaczewski

Institute of Theoretical Physics, Warsaw University Hoża 69, PL-00681, Warsaw, Poland

Physics of nuclei at extreme values of the isospin is at the focus of present-day nuclear science. Experimentally, thanks to existing and emerging radioactive-ion-beam facilities, we are on the verge of invading the territory of extreme $N/Z$ ratios in an unprecedented way. Theoretically, nuclear exotica represent a formidable challenge for the nuclear many-body theories and their power to predict nuclear properties far from stability. Going to the limits of the nuclear binding is also important for an improvement of our description of normal nuclei from the neighborhood of the beta stability valley. In the present talk, we review several aspects of the present-day mean-field theoretical studies of weakly bound nuclei.

## 1. Introduction

In the present paper we attempt a description of issues which pertain to studies of nucleonic system far from stability at low energies. In fact, a number of recent review articles exist already on the subject, see e.g. Refs. [1, 2, 3]. Within a limited space of the present article, we provide only a list of interesting subjects and key questions; this domain of nuclear physics constitutes at present a vast and rapidly growing area of scientific activity. The presentation is focused on the principal goal of studying new aspects of the nuclear effective interaction and nuclear mean field in exotic nuclei. We do not make distinction between interesting theoretical problems *per se* and phenomena and observables in nuclei far from stability. Both aspects are intimately related one to another and can only be addressed simultaneously.

## 2. Effective interactions

Study of the nuclear effective interaction is the main goal of contemporary theory of nucleonic systems. The effective interaction can be, in

(1)

principle, derived from the bare nucleon-nucleon force. Compared to the interaction of two particles in the vacuum, which can be fairly well determined by the scattering experiments, the effective interaction must include the following three physical effects:

- Existence of the hard core, which cannot be treated approximately, but should be described by solving the two-particle problem exactly.

- Modification of the interaction by the medium of nucleons, in which it occurs.

- Reduction in the phase space, used for the description of many-nucleon states.

There is recently a renewed interest in studying properties of effective interactions from first principles [4, 5, 6, 7], and much more effort is still needed. In many practical applications, one uses phenomenological effective interactions fitted to experimental data. Such approaches are tremendously successful, showing that the theoretical concept of the effective interaction very well applies to nucleonic systems. In particular, one may use the effective interaction either in the mean-field approximation or in the shell-model diagonalizations. The former is an approximate, variational method, which has an advantage of being able to use large phase spaces, and in particular the coordinate representation for nucleon wave functions, and can be applied in arbitrarily heavy nuclei. The latter requires strong reduction of the single-particle phase space, but in light nuclei allows for exact solutions, both for the ground and excited states. Strong links which exist at present between the experiment and the approaches using effective interaction call for extension of the observations to nuclei far from stability. This is in fact the most efficient way to establish the properties of effective interactions. Applying the effective interactions in nuclei near the stability valley allows for a consistency check, which then allows to make predictions for nuclei still further out of stability.

## 3. Isovector properties of effective interactions

Effective interactions must depend on numbers of protons and neutrons in systems they aim to describe. This is a consequence of deriving these interactions in the medium and/or in a restricted phase space. However, in practice such dependence can be cast into somewhat more general form, which allows using the same effective interaction in a wide range of nuclei. In particular, the mean-field interactions used at present depend on the local one-body density of the nucleus, and not explicitly on the numbers of particles. Dependence on the isoscalar density is fairly well known, and is dictated by the saturation properties of nuclear forces. On the other hand, dependence on the isovector density is poorly known, and, in fact, is seldom used in applications, because it can manifest itself in a vivid way only when one goes to nuclei with unusual $N/Z$ ratios. Dependence on spin densities, if necessary at all, is entirely unknown at present. However, it requires exploring another class of exotica, namely, the high-spin nuclear states.

The effective interactions used in the shell model implicitly depend on the numbers of particles, because they require choosing the appropriate set of single-particle states in the phase space. Such a construct allows for a description of nuclei between the boundaries of magic particle numbers, or across a given magic particle number, but requires proceeding step-by-step and having experimental feedback when connecting different regions of nuclei. Hence, in this approach the role of experimental data cannot be underestimated. On the other hand, the experimental data at or across the magic numbers may allow for reliable extension of the calculations significantly outside the experimentally accessible region.

## 4. Effective mass

The effective mass is a standard aspect of the mean-field approach and describes to leading order the effects related to the non-locality of the mean field. The non-locality of the mean field may result from a non-locality of the underlying two-body effective interaction, but more importantly, it reflects the Pauli exchange effects in many-fermion systems. In nuclei, the effective mass has the aspect which is characteristic to finite systems, namely, it must depend on the position within the nucleus. Outside the nucleus it approaches the standard free nucleon mass and inside the nucleus it may assume other values. Properties of the effective mass in nuclear systems are not known very well. Its bulk value is usually taken in the range of $m^*=0.6$–$0.8m$, but values outside this range are also used in specific applications. Very little is known about the surface dependence of the effective mass, but more importantly, almost nothing is known about the isovector effective mass, i.e., about the difference between neutron and proton effective masses in the bulk. The effects of the isovector effective mass can only be studied in exotic nuclei, because one needs a substantial difference in proton and neutron numbers and densities in order to amplify these effects to any tangible observables. It is obvious that the densities of single-particle states and the magnitude of shell effects depend on the isovector effective mass, and influence basic properties of nuclei far from stability.

## 5. Spin-orbit interaction

Recently a substantial progress has been made in our understanding of properties of the spin-orbit interaction in nuclei. In principle, the mean-field spin-orbit coupling may result from the Hartree-Fock averaging of the two-body spin-orbit interaction. However, an alternative source is proposed by the relativistic mean field approach [8], where the mean-field spin-orbit coupling stems from the nonrelativistic reduction of the relativistic nucleon spinors. This leads to a specific form of the spin-orbit form factors and their isovector dependence. Needless to say that the isovector properties of the spin-orbit interaction may dramatically influence the shell structure of nuclei far from stability, and therefore, these nuclei can be a perfect testing ground for hypotheses about the source of the spin-orbit coupling in nuclei.

## 6. Pairing near continuum

In weakly bound nuclei, pairing phenomena play a much more important role than in standard systems [9]. Binding of the last unpaired nucleon results from a balance between the mean-field attraction and the missing pairing attraction of the odd nucleon compared to two neighboring even systems. Therefore, in nuclei near drip lines one cannot treat pairing phenomena as a small perturbation but a full coupling of the mean-field and pairing effects is necessary. This brings up a very interesting theoretical aspect of virtual scattering of pairs of nucleons into the phase space of positive single-particle energies. Theoretically, there exist proper tools to describe such pairing phenomena [10], however, experimental tests of our understanding of pairing can only be performed by studying weakly bound systems.

A description of the continuum effects in pairing correlations requires changes in our perception of the positive-energy phase space. Usually, this part of the phase space is accessible through scattering experiments, in which an unbound particle arrives from the infinity and interacts with a bound nucleus, or is emitted by an unbound or excited nucleus and departs to infinity. In studying virtual excitations to positive-energy phase space we do not have any nuclear constituents which would be asymptotically free, and therefore, scattering states are entirely inappropriate elements of the description. On the contrary, one should use a complete set of localized states as a playground where the virtual excitations may take place. Hence, the concept of resonant scattering states is not very useful when describing pairing correlations far from stability.

Our understanding of the effective pairing forces is very far from being complete. Apart from not being able to tell whether the pairing forces are short or long range, we do not know if they are mostly concentrated in the bulk or at the surface, or most importantly, how do they depend on the isovector characteristics of the nuclear medium in the bulk or at the surface.

## 7. Low-density and weakly bound nucleonic matter

Reliable calculations of low-density nuclear and neutron matter exist for infinite medium, see e.g. Ref. [11, 12, 13]. Starting from the bare nucleon-nucleon force, one is able to perform sophisticated theoretical analyses, both perturbatively and variationally, and obtain predictions about binding energy per particle as a function of density. However, these calculations rely heavily on the fact that the system is infinite, and the corresponding calculations in finite nuclei cannot yet be performed. The main question which is not yet answered about the low-density tails of matter distribution in weakly bound systems, is to which extent one can approximate such a region of the nucleus by the infinite matter at a given (low) density. One may reasonably expect that the essential feature of the tail medium is the gradient of the density. Such effects are, of course, entirely beyond any studies performed for the infinite medium. On the other hand, without having any experimental access to infinite (or large) nuclear media, we are bound to study the low-density systems by experimentally investigating properties of the outer tail in weakly bound systems.

## 8. Cluster phenomena in low-density nucleonic matter

The outer tail of the nuclear wave function is a very interesting object, because it can only exist due to the nearby presence of a (nearly) saturated nuclear matter, and otherwise would have been strongly unstable with respect to the clustering effects. Therefore, one can suspect that the clustering correlations may nevertheless play a very important role in the structure of the outer tail of the nuclear density. In light nuclei, the experimental study of the tail may give information about effective interactions in low-density medium, while in heavier systems, which may develop multi-particle low-density tails, one may expect very interesting multi-particle cluster effects to occur.

## 9. One- & two-particle neutron haloes

Weak binding of the last one or two neutrons results in a phenomenon of the neutron halo, see e.g. Ref. [14]. In the halo state, neutrons move at large distances from the center of the nucleus, which gives rise to a significant increase of the reaction radius, and allows investigating structure of the halo by momentum-transfer experiments. Halo states are very well suited for investigating correlations in few-fermion systems. Since the halo neutrons move far away from other nucleons, one can approximate their interaction by the bare nucleon-nucleon force and partly avoid uncertainties related to the construction of the effective interaction.

## 10. Central mean field far from stability

Central-mean-field nuclear potential results from the Hartree-Fock averaging of the central effective interaction with nuclear densities. Independently of the range of the effective interaction (zero or short), nuclear bulk properties depend mostly on the local densities of nucleons, and therefore, we can derive properties of the central mean field directly from the properties of the neutron and proton densities.

Since the strongest part of the effective interaction is the neutron-proton attraction, in stable nuclei the neutron mean field depends mostly on the proton density and *vice versa*. However, in neutron rich nuclei, due to the excess of the neutron density at the surface of the nucleus (Sec. 19), the neutron mean field in the outer region is weak and depends mostly on the neutron distribution, while in the central region is strong and depends mostly on the proton distribution. As a consequence, such a complicated structure of the neutron mean field results in a disappearance of the region where the central field is almost constant.

Therefore, the typical division of the nucleus in the central and surface regions does not hold any more and the neutron mean field gradually decreases from its maximum value near $r=0$, to zero at large distances. This situation is similar to that in light nuclei, where the central region, of an almost-constant mean field, is very narrow, and where one may fairly well approximate the mean filed be the harmonic oscillator potential (apart from its asymptotic properties at large distances, of course). One can reasonably well expect that a similar approximation will also hold in heavy neutron-rich nuclei [15], which may have dramatic consequences on the structure and sequence of the single-particle levels.

## 11. Single-particle energies far from stability

Apart from deformation effects which split single-particle levels in deformed nuclei, the sequence of single-particle states depends mostly on three elements: (i) an overall density of levels and their grouping into major nuclear shells, which depend mostly on the nuclear volume and effective mass, (ii) the order of centroids of spin-orbit partners (they are ordered according to decreasing values of the orbital angular momenta $\ell$), which depends mostly on the flat-bottom region of the nuclear mean field, and (iii) the splitting of the spin-orbit partner levels (high-$j$ partners occur below the low-$j$ partners), which depends on the intensity and the form factor of the spin-orbit mean field.

Hypothetical changes of the neutron mean field in neutron rich nuclei (Sec. 10) may affect the second of these elements. Namely, the disappearance of the flat-bottom region of the central field may decrease the splitting of centroids in such a way that the single-particle spectrum will resemble that of the harmonic oscillator [15, 9]. Moreover, hypothetical modifications of the strength of the spin-orbit coupling in nuclei far from stability (Sec. 5) may affect the third element above. First estimates show [16] that the magnitude of the spin-orbit splitting may decrease in neutron rich nuclei. Changes occurring in the order of single-particle states will be immediately reflected in various measurable properties of nuclei, especially those which are related to collective phenomena (Secs. 14 and 15).

## 12. Proton-neutron pairing

Existence of pairing correlations between like particles is since long a very well established fact in nuclei. In close analogy, one could also expect pairing correlations between neutrons and protons. However, the scalar ($0^+$) proton-neutron pairing may occur only in nuclei with $N \simeq Z$, because only then one has the proton and neutron orbits with the same quantum numbers near the corresponding Fermi levels.

Recently, there is a renewed interest in studying theoretically the proton-neutron correlations, because one obtains more and more experimental information on heavy nuclei with $N \simeq Z$. Collective pairing of this type may only develop in relatively heavy nuclei; otherwise it cannot be well distinguished from a simple proton-neutron interaction of a few valence particles.

Another kind of hypothetical proton-neutron pairing may occur in neutron rich nuclei where the proton and neutron orbits with different parities and the same values of the angular momentum $j$ occur near the Fermi surface. In such a situation, one might expect the pseudoscalar ($0^-$) pairing correlations, which would be a novel phenomenon leading to intrinsic parity-breaking effects of a new kind [17, 18].

## 13. $\alpha$-correlations

Effects related to clustering of nucleons into the alpha particles are very well known in light nuclei. Cluster models work very well for ground and excited states of nuclei like $^8$Be, $^{12}$C, $^{16}$O, $^{20}$Ne, and many others. In heavier $N \simeq Z$ systems one may expect collective $\alpha$-cluster correlations in competition with the proton-neutron pairing correlations. see e.g. Ref. [19]. Although it is not easy to distinguish effects of both kind of correlations, their study may be very fruitful for our understanding on how one should construct the correlated many-body wave functions.

On the other hand, in light neutron-rich nuclei one may expect the $\alpha$-correlations to be hindered by the existence of the neutron skin in which neutrons do not have partner protons for the correlation to develop. Therefore, the $\alpha$-cluster models should be adopted to such a physical situation and correctly predict the isospin dependence of the $\alpha$-correlations.

## 14. Deformations of weakly-bound systems

Study of the deformability of weakly bound nuclear systems constitutes at present one of the most important challenges of the nuclear structure physics. Indeed, a proper description requires taking into account the pairing, continuum, and deformation effects at the same time. Closeness of the particle continuum may increase the effective density of single-particle states as compared to stable nuclei, which can be due to the presence of the positive-energy phase space near the Fermi level, Sec. 6. This may decrease the deformability of nuclei, because the chance to find a large shell gap at a given deformation will be smaller. Moreover, an increase of density of states may increase the intensity of pairing correlations, and further hinder the deformation effects. However, we may also encounter situations in which the Fermi energy is negative only at deformed shapes, and positive at sphericity, and then deformation will be the only way for a given nucleus to be stable.

Modifications of the deformation patterns in nuclei far from stability can provide the simplest signals of the corresponding changes in the single-particle shell structure. In stable nuclei, deformations follow simple and well recognized patterns, and a departure form the pattern can be attributed to modifications of the intrinsic structure [20]. In fact, if the pattern becomes modified, we have a strong argument for an expected changed structure, while if it does not, we have a fascinating theoretical puzzle to solve.

In very neutron rich deformed nuclei one may also find interesting structures of the neutron haloes. Namely, a deformed single-particle orbital must asymptotically have a spherically symmetric tail wave functions, and the transition between both regions in space may lead to very interesting observable effects. see the recent analysis in Ref. [21].

## 15. Collectivity in weakly-bound systems

Structure of collective excitations is defined by the energies of the elementary single-particle excitations (shell gaps) and the relation between spatial properties of the excitation modes and wave functions. The latter dependence may lead in neutron rich nuclei to significant modifications of the collective modes. Due to a decoupling of neutrons and protons in the surface region, one may expect low-lying isovector modes of various multipolarities, for example, dipole vibrations of the proton core with respect to the neutron skin or quadrupole vibrations of a deformed proton core with respect to spherical or deformed neutron skin. Intensities of such modes are direct consequences of the effective interactions in the isovector channels. In stable nuclei, isovector excitations are difficult to excite and measure, because they often lead to high excitation energies and mix with single-particle structures. However, in very neutron rich nuclei they may become the lowest excitations available to the system.

## 16. One-proton emitters

Significant progress has recently been achieved in studying nuclei at and beyond the proton drip line [22]. Odd-proton nuclei beyond the proton drip line may decay by emitting protons, which gives us invaluable information on positions of proton quasi-bound states within the Coulomb barrier. These states are in fact extremely narrow resonances, which for all practical purposes behave like bound states. In heavy nuclei, the Coulomb barrier is high, and therefore the energies of the quasi-bound states follow the standard sequence of the bound states in the average proton mean field. The barrier-penetration probability depends mainly on the shape of the outer barrier, which is given by pure Coulomb potential, and on the quantum numbers of the quasi-bound proton state. Therefore, study of proton half lives gives precise information on the nuclear mean field at the surface, and in particular about the nuclear deformation, see e.g. Ref. [23]. Fine structure studies of the proton spectra may reveal isomeric states in the proton unstable nucleus and excited states in the daughter nucleus.

## 17. Two-proton emitters

Phenomenon of a single-step two-proton decay still awaits its discovery, apart from such a process being known already in very light nucleus $^6$Be. Several candidates for such a decay have been singled out by theoretical estimations of the decay energies and half lives [24, 25]. However, competition between the two-proton decay and the $\beta$ decay depends extremely strongly on the single-particle proton energies, and therefore, experimental study of such a process can only be possible in a few very specific cases. Nevertheless, measurement of the relative momenta of the emitted protons can provide invaluable information on the proton correlations in the parent nucleus, and on the properties of the final-state interaction.

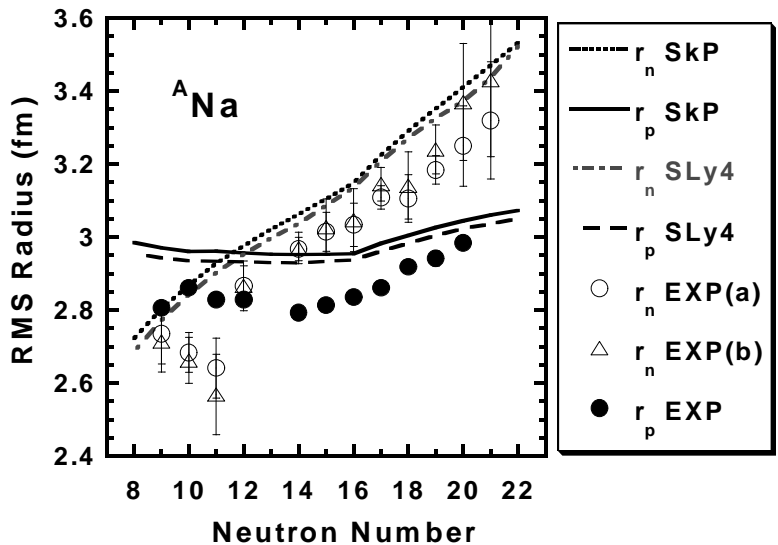

Fig. 1. Neutron and proton radii in sodium isotopes. Experimental data [26] are compared with the spherical HFB calculations performed with the SkP [10] and SLy4 [27] Skyrme interactions.

## 18. Neutron radii

Direct measurement of neutron radii are not easy, because the electromagnetic probes (electrons) are mostly sensitive to the distribution of charge, while the hadronic probes (protons, antiprotons and mesons) do not well distinguish neutrons and protons in the nucleus. Nevertheless, in several cases the neutron root-mean-squared radii could be extracted from the reaction cross sections, and agree fairly well with the mean-field calculations, both in stable nuclei, see Ref. [28] and references cited therein, and in neutron-rich nuclei [26, 29, 30, 31, 32]. This includes predictions of the thickness of the neutron skin, namely, the difference between the neutron and proton radii, which gradually increase with the neutron excess. As an example of the agreement, we show in Fig. 1 results of the Hartree-Fock-Bogoliubov (HFB) calculations compared with experimental data [26]. Except from a slight overall overestimation of the radii, one very easily obtains proper qualitative reproduction of the difference between neutron and proton radii. At present, experimental errors of the neutron radii are significant, which precludes obtaining from these data detailed spectroscopic informations, such as we have at our disposal for the charge radii.

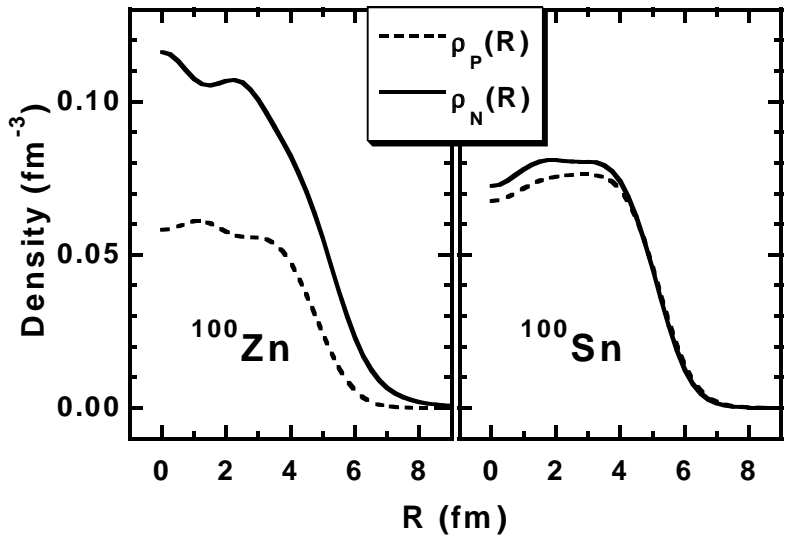

Fig. 2. Neutron and proton density distributions in $^{100}$Zn and $^{100}$Sn calculated within the HFB theory with the SkP Skyrme interaction.

## 19. Density distributions far from stability

Apart form the fact that a long tail of the matter distribution may be present in nuclei near drip lines, the profiles of densities in the central and surface regions of such a nuclei may be modified too. In nuclei with $N=Z$, density distributions of protons and neutrons are fairly similar, see results for $^{100}$Sn presented in Fig. 2. This results from a strong attraction between protons and neutrons, which equalizes densities of protons and neutrons both in the central and surface regions. Compared to nuclear forces, the Coulomb interaction is fairly weak, and is only able to slightly shift the proton density from the central to surface region.

On the other hand, in nuclei with a significant excess of neutrons, $N \gg Z$, such as $^{100}$Zn illustrated in Fig. 2, the neutron and proton density distributions must be different just because of the different numbers of the two kind of nucleons. In discussing such differences one may single out two opposite and extreme scenarios. Namely, (i) the neutron distribution may assume the same shape as the proton distribution, and be just a factor of $N/Z$ larger. In this case, the neutron and proton surface regions coincide, while the corresponding central densities differ very much, which is in opposition to the strong attraction between neutrons and protons in the bulk. On the other side, (ii) the neutron density in the central region may be equal to that of protons, but then it should extend much further out beyond the proton distribution, which is in opposition to strong surface attraction between neutrons and protons. Realistic neutron distributions will of course be a compromise between these to opposite scenarios, and the frustration of neutrons, not knowing where it is better to go, will result in differences of neutron and proton densities both in the central *and* in the surface region of a nucleus.

In the mean field approaches based on the local density approximation, the bulk and surface neutron-proton attractions are governed by terms of a different structure, namely, $(\rho_n - \rho_p)^2$ and $(\vec{\nabla}\rho_n - \vec{\nabla}\rho_p)^2$, respectively. Coupling constants, which give the strength of these two terms, are adjusted to global nuclear properties, like the symmetry energy and surface symmetry energy, and therefore, the balance between the bulk and surface attraction is fairly well defined. In nuclei far from stability, the mean-field calculations based on such derivation, produce neutron distributions which significantly differ from those for protons, *both* in the central *and* surface regions. As a simple consequence of this fact one obtains in neutron rich nuclei an increased surface diffuseness of the neutron distribution. [15].

This is illustrated in Fig. 3, where the neutron surface thickness $a$ has been derived from the neutron distributions calculated within the spherical HFB approach [10] with the SkP [27] Skyrme interaction. The simplest method of derivation has been used, namely, the microscopic densities have been fitted by the Fermi distribution

$$\rho(R) = \frac{\rho_0}{1 + \exp\left(\frac{R-R_0}{a}\right)}, \qquad (1)$$

where the central density $\rho_0$, equivalent radius $R_0$, and surface diffuseness $a$ are the adjustable parameters. On can see that a lot of structure information appears in the surface diffuseness of neutron distributions. Apart from a gradual increase with neutron excess, one can clearly see well pronounced shell effects related to closure of major neutron shells. Therefore, even crude measurements of the neutron diffuseness may provide invaluable information about shell structure of neutron-rich nuclei. This can be opposed with measurements of neutron radii (Sec. 18), which need to be measured fairly precisely in order to provide information on the underlying shell structure.

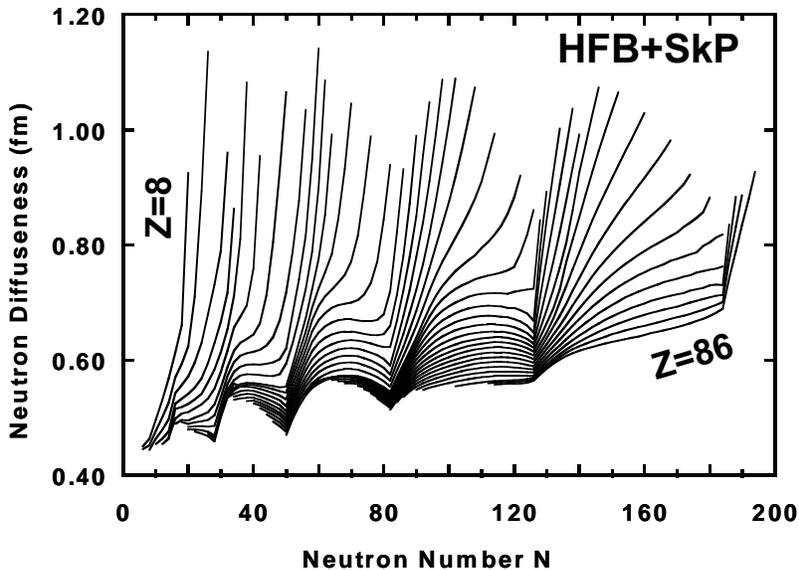

Fig. 3. Neutron diffuseness calculated from the HFB neutron distributions in even-even nuclei. Microscopic distributions have been approximated by simple Fermi shapes, Eq. (1), and the values of the obtained parameters $a$ are plotted here as functions of the neutron number. Lines connect results for nuclei with given numbers of protons, from $Z=8$ to $Z=86$.

## 20. Conclusion

Experimental and theoretical studies of nuclei far from stability become the main subjects of the present-day nuclear physics. Long-range plans and proposal [33, 34, 35, 36], in Europe, Japan, and United States call for an increased activity in this field and point out enormous scientific interest of investigating systems with extreme $N/Z$ ratios. In the present short account of prospective scientific opportunities, we were merely able to enumerate various aspects of the physics involved. Undoubtedly, we can expect in the near future many fascinating new results to emerge, along with an unforeseeable number of unexpected discoveries.

## ACKNOWLEDGMENTS

This research was supported in part by the Polish Committee for Scientific Research (KBN) under Contract No. 2 P03B 040 14.